# State-Dependent Multiple Access Channel with Feedback

Saeed Hajizadeh

(Undergraduate Student)
Department of Electrical Engineering
Ferdowsi University of Mashhad
Mashhad, Iran
Saeed.hajizadeh1367@gmail.com

Ghosheh Abed Hodtani

Department of Electrical Engineering
Ferdowsi University of Mashhad
Mashhad, Iran
ghodtani@gmail.com

*Abstract—* **In this paper, we examine discrete memoryless Multiple Access Channels (MACs) with two-sided feedback in the presence of two correlated channel states that are correlated in the sense of Slepian-Wolf (SW). We find achievable rate region for this channel when the states are provided non-causally to the transmitters and show that our achievable rate region subsumes Cover-Leung's achievable rate for the discrete memoryless MAC with two-sided feedback as its special case. We also find the capacity region of discrete memoryless MAC with two-sided feedback and with SW-type correlated states available causally or strictly causally to the transmitters. We also study discrete memoryless MAC with partial feedback in the presence of two SW-type correlated channel states that are provided non-causally, causally, or strictly causally to the transmitters. An achievable rate region is found when channel states are non-causally provided to the transmitters whereas capacity regions are characterized when channel states are causally, or strictly causally available at the transmitters.**

## I. Introduction

MAC is one of the most scrutinized multi-user channels and consists of two transmitters and a receiver. The two transmitters try to reliably send their own private messages to the receiver. MAC was first studied in [1] and [2] where the capacity region of the MAC with independent sources was found. Slepian and Wolf [3] found the capacity region of the discrete memoryless two-user MAC with correlated sources. Cover, El Gamal, and Salehi [4] discussed MAC with arbitrarily correlated sources and provided an achievable rate region which contained those of [1], [2], and [3] as its special cases. Gaarder and Wolf [5] have shown through the use of a binary example that the capacity region of MAC can be increased using noiseless feedback from the receiver to both the transmitters. Cover and Leung [6] later found an achievable rate region for the MAC with feedback using the notion of *lexicographical indexing* and *list decoding*.

Willems [7] later proved the optimality of [6] in a special case where at least one of the inputs is a function of the output and the other input. Willems and Van Der Meulen [8] proved that the achievable rate region provided in [6] is also achievable for the MAC with partial feedback. Channels with SI were first studied by Shannon [9] where he found the capacity of the single-user channel with SI causally available at the transmitter. The capacity of the single-user channel with SI available non-causally at the transmitter was determined by Gelf'and and Pinsker [10]. Multiple access channels with SI have been studied in [11]-[17].

In this paper,

1. We first study discrete memoryless MAC with two-sided noiseless feedback with two correlated channel states in the sense of Slepian-Wolf. We provide an achievable rate region for this channel when channel states are provided non-causally to the transmitters and we show that our achievable rate region for this channel subsumes Cover-Leung's achievable rate. We also characterize the capacity region of this channel in two cases, i.e. when the channel states are provided causally and when the channel states are provide strictly causally to the transmitters.

2. We then scrutinize the discrete memoryless MAC with one-sided (partial) feedback with two channel states that are correlated in the sense of Slepian-Wolf. We derive an achievable rate region for this channel when channel states are provided non-causally at the transmitters. We also provide the capacity region of this channel when states are provided causally or strictly causally to the senders.

By SW-type correlated SI available at the transmitters of a discrete memoryless MAC we mean that if, for instance, the two correlated states available at senders 1 and 2 are $\tilde{S}_1$ and $\tilde{S}_2$, respectively, we can model these states with three pair-wise independent states $S_0, S_1$, and $S_2$ where $\tilde{S}_1 = (S_0, S_1)$, and $\tilde{S}_1 = (S_0, S_1)$, and $S_0$ is available to both the transmitters and $S_k$ is only available at transmitter $k, k = 1,2$.

The rest of the paper is organized as follows. In section II, definitions are given. In section III, the main results of the paper are provided while section IV is devoted to the conclusion.

## II. DEFINITIONS

A discrete Memoryless MAC with feedback with correlated SI non-causally available at the transmitter is depicted in Fig.1.

*Definition 1:* The discrete memoryless MAC with SW-type correlated SI non-causally available at the transmitters and with feedback denoted by $(\mathcal{X}_1, \mathcal{X}_2, \mathcal{Y}, \mathcal{S}_0, \mathcal{S}_1, \mathcal{S}_2, p(y|x_1, x_2, s_0, s_1, s_2))$ consists of three finite sets $\mathcal{X}_1, \mathcal{X}_2$, and $\mathcal{Y}$ and a collection of probability mass functions $p(y|x_1, x_2, s_0, s_1, s_2)$ on $\mathcal{Y}$, one for each $(x_1, x_2, s_0, s_1, s_2) \in \mathcal{X}_1 \times \mathcal{X}_2 \times \mathcal{S}_0 \times \mathcal{S}_1 \times \mathcal{S}_2$. The channel is memoryless, therefore:

$$p(y^n|x_1^n, x_2^n, s_0^n, s_1^n, s_2^n) = \prod_{i=1}^{n} p(y_i|x_{1i}, x_{2i}, s_{0i}, s_{1i}, s_{2i})$$

A $((2^{nR_1}, 2^{nR_2}), n)$ code for the discrete memoryless MAC with SI non-causally available at the transmitters and with feedback consists of two collections of $n$ encoder mappings

$x_{1i}: \{1,2, \ldots, M_1\} \times \mathcal{Y}^{i-1} \times \mathcal{S}_0^n \times \mathcal{S}_1^n \to \mathcal{X}_1$, $i = 1,2, \ldots, n$

$x_{2i}: \{1,2, \ldots, M_2\} \times \mathcal{Y}^{i-1} \times \mathcal{S}_0^n \times \mathcal{S}_2^n \to \mathcal{X}_2$, $i = 1,2, \ldots, n$

and a decoding function

$$g: \mathcal{Y}^n \to \{1,2, \ldots, M_1\} \times \{1,2, \ldots, M_2\}$$

Therefore each sender makes use of the message $m_k$, the past symbols of the output $y_1, y_2, \ldots, y_{i-1}$, and the whole sequence of $s_0^n$, and $s_k^n$ to produce $x_{k,i}$, i.e. the ith component of $x_k^n, k = 1,2$.

Assuming that each message is distributed uniformly and independently on its respective set, we have

$$P_e^{(n)} = \Pr\{g(Y^n) \neq (M_1, M_2)\}$$

$$= \frac{1}{M_1 M_2} \sum_{m_1, m_2} p(g(y^n) \neq (m_1, m_2)|m_1, m_2 \text{ sent})$$

A rate pair $(R_1, R_2)$ is achievable for the channel in Fig. 1 if there exists a $((2^{nR_1}, 2^{nR_2}), n)$ code with $P_e^{(n)} \to 0$ as $n \to 0$. The capacity region is the closure of the convex hull of the set of all achievable rates.

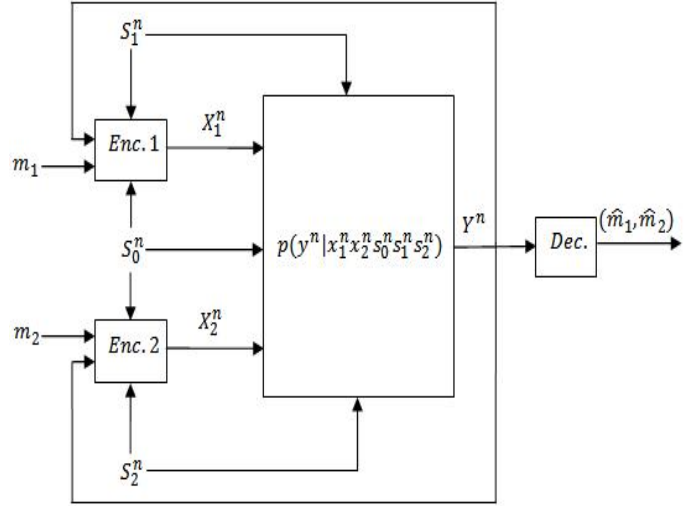

Figure 1. Multiple Access Channel with Correlated Side Information non-causally available at the transmitters and with feedback.

*Definition 2:* The discrete memoryless MAC with correlated SI causally available at the transmitters and with feedback, depicted in Fig. 2, has the same channel and achievable rate definitions as that of Fig. 1 except a slight difference in the code used. In this case, there are two collections of $n$ encoder mappings

$x_{1i}: \{1,2, \ldots, M_1\} \times \mathcal{Y}^{i-1} \times \mathcal{S}_0^i \times \mathcal{S}_1^i \to \mathcal{X}_1$, $i = 1,2, \ldots, n$

$x_{2i}: \{1,2, \ldots, M_2\} \times \mathcal{Y}^{i-1} \times \mathcal{S}_0^i \times \mathcal{S}_2^i \to \mathcal{X}_2$, $i = 1,2, \ldots, n$

*Definition 3:* A discrete memoryless MAC with feedback and with correlated SI strictly causally available at the transmitters, depicted in Fig. 3, is defined just as its counterpart with causal SI except that the code definition is slightly different, i.e. there exists an integer $0 < r \le i$ such that

$x_{1i}: \{1,2, \ldots, M_1\} \times \mathcal{Y}^{i-1} \times \mathcal{S}_0^{i-r} \times \mathcal{S}_1^{i-r} \to \mathcal{X}_1$, $i = 1,2, \ldots, n$

$x_{2i}: \{1,2, \ldots, M_2\} \times \mathcal{Y}^{i-1} \times \mathcal{S}_0^{i-r} \times \mathcal{S}_2^{i-r} \to \mathcal{X}_2$, $i = 1,2, \ldots, n$

*Definition 4:* The discrete memoryless MAC with partial feedback and with correlated SI non-causally available at the transmitters is depicted in Fig. 4. The channel, the achievable rate region definitions, and part of the code remain the same as those of definition 1. The encoder mappings for the second transmitter, though, alter so that it reflects the difference between partial and full feedback, i.e. we have two sets of encoding functions as follows

$x_{1i}: \{1,2, \ldots, M_1\} \times \mathcal{Y}^{i-1} \times \mathcal{S}_0^n \times \mathcal{S}_1^n \to \mathcal{X}_1$, $i = 1,2, \ldots, n$

$x_{2i}: \{1,2, \ldots, M_2\} \times \mathcal{S}_0^n \times \mathcal{S}_2^n \to \mathcal{X}_2$, $i = 1,2, \ldots, n$

Notice that the discrete memoryless MAC with partial feedback and with correlated SI causally or strictly causally available at the transmitters is defined with the negligible variations in the definition of the code.

## III. MAIN RESULTS

Define $\mathcal{P}_1^*$ as the set of all distributions on the collection of all random variables $(S_0, S_1, S_2, U, V_1, V_2, X_1, X_2, Y)$ with finite alphabets such that

$$p(s_0, s_1, s_2, u, v_1, v_2, x_1, x_2, y) =$$
$$p(s_0)p(s_1)p(s_2)p(u)p(v_1|u, s_0, s_1)p(v_2|u, s_0, s_2) \times$$
$$p(x_1|u, v_1, s_0, s_1)p(x_2|u, v_2, s_0, s_2)p(y|x_1, x_2, s_0, s_1, s_2) \quad (1)$$

where

$$p(x_k|u, v_k, s_0, s_k) \in \{0,1\}, \quad k = 1,2$$

Define $\mathcal{R}_{MACTSF-NCSI}(p)$ as the set of all rate pairs $(R_1, R_2)$ satisfying

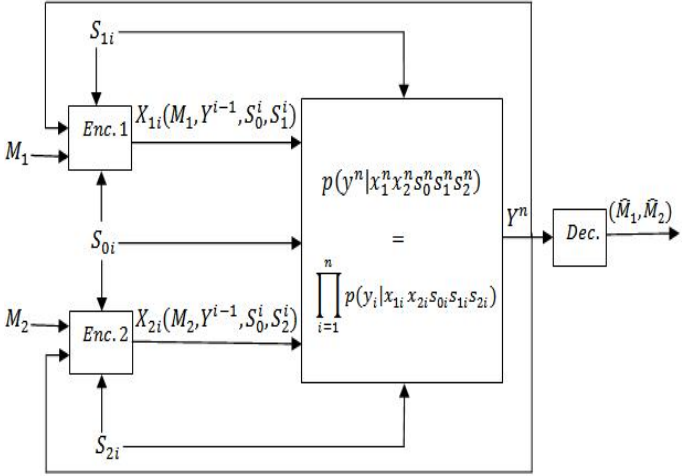

Figure 2. Multiple Access Channel with Correlated Side Information Causally available at the transmitters and with feedback.

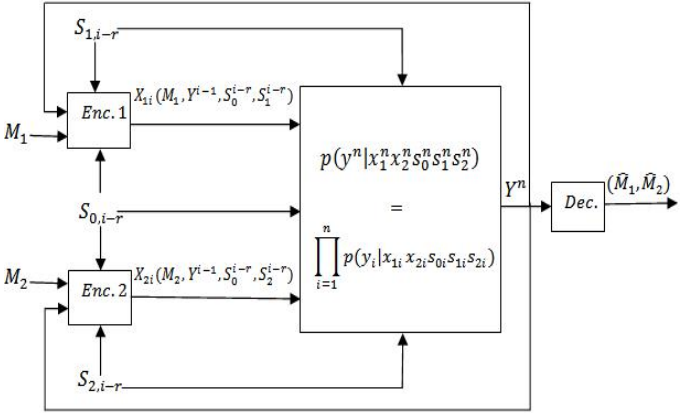

Figure 3. Multiple Access Channel with Correlated Side Information strictly causally available at the transmitters and with feedback.

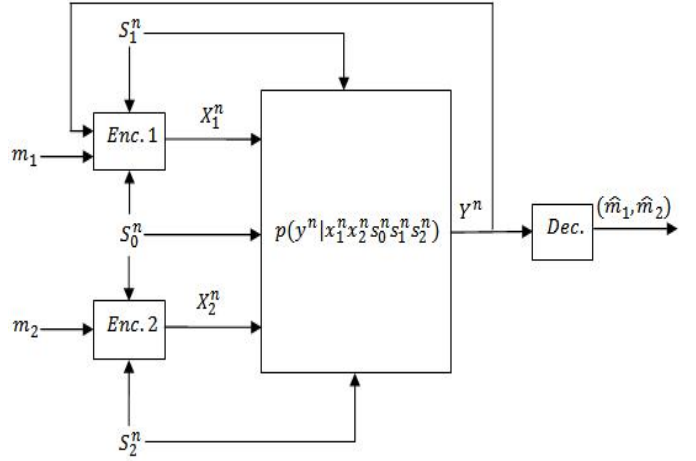

Figure 4. Multiple Access Channel with partial feedback and with Correlated Side Information.

$$\begin{aligned}
R_1 &\leq \min\{I(V_1;Y|UV_2), I(V_1;Y|UV_2S_0S_2)\} \\
&\quad - I(V_1; S_0S_1|U) \\
R_2 &\leq \min\{I(V_2;Y|UV_1), I(V_2;Y|UV_1S_0S_1)\} \\
&\quad - I(V_2; S_0S_2|U)
\end{aligned} \quad (2)$$

$$R_1 + R_2 \leq I(V_1V_2;Y) - I(V_1; S_0S_1|U) - I(V_2; S_0S_2|U)$$

for some $p \in \mathcal{P}_1^*$.

***Theorem 1:*** The set $\mathcal{R}^*_{MACTSF-NCSI}$ is achievable for the discrete memoryless MAC with two-sided feedback and with SI non-causally available at the transmitters

$$\mathcal{R}^*_{MACTSF-NCSI} = cl\left(co\left(\bigcup_{p \in \mathcal{P}_1^*} \mathcal{R}_{MACTSF-NCSI}(p)\right)\right) \quad (3)$$

where $cl(co(.))$ is the closure of the convex hull operation.

***Remark 1.1:*** If we set $S_0 \equiv S_1 \equiv S_2 \equiv \emptyset$ in (2) we derive Cover-Leung achievable rate region [6] for MAC with feedback.

***Proof:*** The proof uses the notions of superposition block Markov, and Gelf'and-Pinsker encoding. We use a large number $B$ of blocks to convey $B-1$ messages $(m_{k1}, m_{k2}, \dots, m_{kB-1})$, $k = 1,2$, to the receiver with arbitrarily small probability of error. In each block $b$, the transmitters send two types of information. They first cooperate through the noiseless feedback links to decode each other's messages sent to the receiver in the previous block, i.e. block $b-1$, and choose the right index to send to the receiver so that the receiver is able to resolve its remaining uncertainty about the messages left over from the previous block. Notice that at the end of each block, the receiver can only partially decode that block's messages, i.e. the messages sent at the beginning of block $b$ are partially decoded at the end of block $b$ and completely decoded at the end of block $b+1$. The transmitters also superimpose new information on the index chosen at the end of each block. This plan leads to a steady-state infusion of new information and resolution of the remaining uncertainty. We now stick to the coding:

***Codebook Generation:*** Assume that
$$p(s_0^n, s_1^n, s_2^n) = \prod_{i=1}^{n} p(s_{0i}) p(s_{1i}) p(s_{2i})$$

Fix a joint distribution $p(u)p(v_1|u, s_0, s_1)p(v_2|u, s_0, s_2)$. Randomly and independently generate $2^{nR_0}$ n-sequences of $u^n(m_0)$ each one i.i.d according to $\prod_{i=1}^{n} p(u_i)$. For each $u^n(m_0)$, randomly and conditionally independently generate $2^{n(R_1'+R_1)}$ n-sequences $v_1^n(m_0, m_1', m_1)$ each one i.i.d according to $\prod_{i=1}^{n} p(v_{1i}(m_0, m_1', m_1)|u_i(m_0))$ and randomly pour them into $2^{nR_1}$ bins. For each $u^n(m_0)$, randomly and conditionally independently generate $2^{n(R_2'+R_2)}$ n-sequences $v_2^n(m_0, m_2', m_2)$ each one i.i.d according to $\prod_{i=1}^{n} p(v_{2i}(m_0, m_2', m_2)|u_i(m_0))$ and randomly pour them into $2^{nR_2}$ bins. Reveal the codebook to both the transmitters and the receiver. As it is described in the coding scheme, the receiver can decode the cloud center $u^n(m_0)$ at the end of each block in which the cloud is sent. While the satellite indices, i.e. the messages $m_1$ and $m_2$, are completely decoded by senders 2 and 1, respectively, the receiver can only partially decode them at the end of the block in which they are sent. Notice that in the first block we have no helping index $m_0$, therefore we superimpose the messages intended for the receiver in the first block on a randomly chosen index. In the last block, though, the receiver receives enough information to resolve its remaining uncertainty about the messages sent in the penultimate block.

***Encoding:*** Assume that the helping index we wish to send to the receiver in block $b$ to resolve its remaining uncertainty about the messages in block $b-1$ is $m_{0b}$. Also suppose that the new messages we would like to send to the receiver in block $b$ are $m_{1b}$ and $m_{2b}$. Both senders choose $u^n(m_{0b})$. Then, the first transmitter searches in bin $m_{1b}$ and look for some $m_{1b}'$ such that
$$(u^n(m_{0b}), v_1^n(m_{0b}, m_{1b}', m_{1b}), s_0^n, s_1^n) \in A_\epsilon^{(n)}$$
where $A_\epsilon^{(n)}$ is the set of $\epsilon$-typical sequences.

The second transmitter, meanwhile, searches the bin $m_{2b}$ and looks for some $m_{2b}'$ such that
$$(u^n(m_{0b}), v_2^n(m_{0b}, m_{2b}', m_{2b}), s_0^n, s_2^n) \in A_\epsilon^{(n)}$$

Suppose that the true binning indices chosen are $m_{kb}' = M_{kb}'$, $k = 1,2$. Then the sequences $x_1^n$ and $x_2^n$ where
$$x_{ki} = f_k(u_i(m_0), v_{ki}(m_{0b}, M_{kb}', m_{kb}), s_0^n, s_k^n) \quad k = 1,2$$
are sent through the channel.

***Decoding:*** The decoder receives $y^n(b)$ at the end of block $b$ and declares $\hat{m}_{0b} = m_{0b}$ as the index sent iff there is a unique $m_{0b}$ such that
$$(u^n(m_{0b}), y^n(b)) \in A_\epsilon^{(n)}$$

Transmitter 1 declares $\hat{m}_{2b} = m_{2b}$ iff there exists a unique $m_{2b}$ such that
$$(u^n(m_{0b}), v_1^n(m_{0b}, M_{1b}', m_{1b}), v_2^n(m_{0b}, M_{2b}', m_{2b})$$
$$, y^n(b), s_0^n, s_1^n) \in A_\epsilon^{(n)}$$

Transmitter 2 declares $\hat{m}_{1b} = m_{1b}$ iff there exists a unique $m_{1b}$ such that
$$(u^n(m_{0b}), v_1^n(m_{0b}, M_{1b}', m_{1b}), v_2^n(m_{0b}, M_{2b}', m_{2b})$$
$$, y^n(b), s_0^n, s_2^n) \in A_\epsilon^{(n)}$$

Now consider the set $\mathcal{T}_{y^n}$ of all codewords that are jointly typical with $y^n$. Define $\xi_{m_1 m_2}(y^n)$ to equal 1 if
$$(u^n(m_0), v_1^n(m_0, M_1', m_1), v_2^n(m_0, M_2', m_2), y^n) \in A_\epsilon^{(n)}$$
and equal 0 otherwise. Now we bound the cardinality of the set $\mathcal{T}_{y^n}$, we have
$$\| \mathcal{T}_{y^n} \| = \sum_{m_1, m_2} \xi_{m_1 m_2}(y^n) \quad (4)$$

Taking expectation from both sides of the above equation we have
$$E \| \mathcal{T}_{y^n} \| = \sum_{m_1, m_2} E\left(\xi_{m_1 m_2}(y^n)\right) = E\left(\xi_{m_{1b} m_{2b}}(y^n)\right)$$
$$+ \sum_{\substack{m_1 \neq m_{1b} \\ m_{1b}' \neq M_{1b}'}} E\left(\xi_{m_1 m_2}(y^n)\right) + \sum_{\substack{m_2 \neq m_{2b} \\ m_{2b}' \neq M_{2b}'}} E\left(\xi_{m_1 m_2}(y^n)\right)$$
$$+ \sum_{\substack{m_1 \neq m_{1b} \\ m_{1b}' \neq M_{1b}'}} \sum_{\substack{m_2 \neq m_{2b} \\ m_{2b}' \neq M_{2b}'}} E\left(\xi_{m_1 m_2}(y^n)\right)$$
$$\leq 1 + \left(2^{n(R_1'+R_1)} - 1\right) 2^{-n(I(V_1;Y|UV_2)-\epsilon)}$$
$$+ \left(2^{n(R_2'+R_2)} - 1\right) 2^{-n(I(V_2;Y|UV_1)-\epsilon)}$$
$$+ \left(2^{n(R_1'+R_1)} - 1\right)\left(2^{n(R_2'+R_2)} - 1\right) 2^{-n(I(V_1 V_2;Y|U)-\epsilon)} \quad (5)$$

Now if
$$R_1' + R_1 \leq I(V_1; Y|UV_2) - \epsilon \quad (6)$$
$$R_2' + R_2 \leq I(V_2; Y|UV_1) - \epsilon \quad (7)$$
then we have
$$E \| \mathcal{T}_{y^n} \| \leq 1 + \epsilon + 2^{n(\Psi+\epsilon)} \leq 2^{n(\Psi+\epsilon')} \quad (8)$$

where $\Psi \triangleq R_1' + R_1 + R_2' + R_2 - I(V_1V_2; Y|U)$ and $\epsilon' > 0$.

Now using Markov's inequality we have

$$pr\{\| \mathcal{T}_{y^n} \| \geq 2^{n\epsilon''} 2^{n(\Psi+\epsilon')}\} \leq \frac{E \| \mathcal{T}_{y^n} \|}{2^{n\epsilon''} 2^{n(\Psi+\epsilon')}} \leq 2^{-n\epsilon''} \leq \epsilon \quad (9)$$

where $\epsilon'' > 0$ and in the penultimate inequality we have used (8).

*Analysis of the error probability:* The potential error events in the encoding or decoding stages during block b are as follows

$$E_1^{en} = \{(u^n(m_{0b}), v_1^n(m_{0b}, m_{1b}', m_{1b}), s_0^n, s_1^n) \notin A_\epsilon^{(n)}$$
$$for\ all\ m_{1b}' \in \{1,2,\dots,2^{nR_1'}\}\}$$

$$E_2^{en} = \{(u^n(m_{0b}), v_2^n(m_{0b}, m_{2b}', m_{2b}), s_0^n, s_2^n) \notin A_\epsilon^{(n)}$$
$$for\ all\ m_{2b}' \in \{1,2,\dots,2^{nR_2'}\}\}$$

$$E_{11}^{dec} = \{(u^n(m_{0b}), y^n(b)) \notin A_\epsilon^{(n)}\}$$

$$E_{12}^{dec} = \{(u^n(\hat{m}_{0b}), y^n(b)) \in A_\epsilon^{(n)}\ for\ some\ \hat{m}_{0b} \neq m_{0b}\}$$

$$E_{21}^{dec} = \{(u^n(m_{0b}), v_1^n(m_{0b}, M_{1b}', m_{1b}), v_2^n(m_{0b}, M_{2b}', m_{2b})$$
$$, y^n(b), s_0^n, s_1^n) \notin A_\epsilon^{(n)}\}$$

$$E_{22}^{dec} = \{(u^n(m_{0b}), v_1^n(m_{0b}, \hat{m}_{1b}', \hat{m}_{1b}), v_2^n(m_{0b}, M_{2b}', m_{2b})$$
$$, y^n(b), s_0^n, s_1^n) \in A_\epsilon^{(n)}\ for\ some\ \hat{m}_{1b}' \neq M_{1b}'\ and\ \hat{m}_{1b} \neq m_{1b}\}$$

$$E_{31}^{dec} = \{(u^n(m_{0b}), v_1^n(m_{0b}, M_{1b}', m_{1b}), v_2^n(m_{0b}, M_{2b}', m_{2b})$$
$$, y^n(b), s_0^n, s_2^n) \notin A_\epsilon^{(n)}\}$$

$$E_{32}^{dec} = \{(u^n(m_{0b}), v_1^n(m_{0b}, M_{1b}', m_{1b}), v_2^n(m_{0b}, \hat{m}_{2b}', \hat{m}_{2b})$$
$$, y^n(b), s_0^n, s_2^n) \in A_\epsilon^{(n)}\ for\ some\ \hat{m}_{2b}' \neq M_{2b}'\ and\ \hat{m}_{2b} \neq m_{2b}\}$$

Now using the standard error probability calculation methods we see that the error probability in block $b$ is small provided that

$$R_1' \geq I(V_1; S_0S_1|U) \quad (10)$$
$$R_2' \geq I(V_2; S_0S_2|U) \quad (11)$$
$$R_0 \leq I(U; Y) \quad (12)$$
$$R_1' + R_1 \leq I(V_1; Y|UV_2S_0S_2) \quad (13)$$
$$R_2' + R_2 \leq I(V_2; Y|UV_1S_0S_1) \quad (14)$$

Finally due to (9) the helping index takes on no more than $2^{n(\Psi+\epsilon')}$ values, therefore using (12) we have

$$R_1' + R_1 + R_2' + R_2 - I(V_1V_2; Y|U) + \epsilon' \leq I(U; Y)$$

Thus

$$R_1' + R_1 + R_2' + R_2 \leq I(V_1V_2; Y) - \epsilon' \quad (15)$$

Now combining (6), (7), (13), (14), (15) with (10) and (11),

the rates in (2) are derived and therefore the proof of the Theorem is complete ∎

*Remark 1.2:* If we remove the feedback links, then each transmitter cannot decode the other transmitter's private message and therefore rate splitting is not needed and we can set in (2), $M_{0k} \equiv \emptyset$ and $M_k \equiv M_{kk}, k = 1,2$, and $U \equiv \emptyset$. Therefore we have the following rate region

$$\begin{cases} R_1 \leq I(V_1; Y|V_2) - I(V_1; S_0S_1) \\ R_1 \leq I(V_2; Y|V_1) - I(V_2; S_0S_2) \\ R_1 + R_2 \leq I(V_1V_2; Y) - I(V_1; S_0S_1) - I(V_2; S_0S_2) \\ for\ some\ p(s_0, s_1, s_2, v_1, v_2, x_1, x_2, y) = \\ p(s_0)p(s_1)p(s_2)p(v_1, x_1|s_0, s_1)p(v_2, x_2|s_0, s_2) \times \\ p(y|x_1, x_2, s_0, s_1, s_2) \end{cases}$$

Now if we set $S_0 \equiv S_1 \equiv S_2 \equiv S$ in the above region, we derive the achievable rate region for the state-dependent MAC with independent sources, i.e. the state-dependent version of [2].

Define $C_{MACTSF-CSI}(p)$ as the set of all rate pairs $(R_1, R_2)$ satisfying

$$R_1 \leq \min\{I(V_1; Y|UV_2), I(V_1; Y|UV_2S_0S_2)\}$$
$$R_2 \leq \min\{I(V_2; Y|UV_1), I(V_2; Y|UV_1S_0S_1)\} \quad (16)$$
$$R_1 + R_2 \leq I(V_1V_2; Y)$$

for some $p \in \mathcal{P}_1^*$.

*Theorem 2:* The set $C_{MACTSF-CSI}^*$ is the capacity region of the discrete memoryless MAC with two-sided feedback and with SW-type correlated SI causally available at the transmitters

$$C_{MACTSF-CSI}^* = cl\left(co\left(\bigcup_{p \in \mathcal{P}_1^*} C_{MACTSF-CSI}(p)\right)\right) \quad (17)$$

*Remark 2.1:* If we set $S_0 \equiv S_1 \equiv S_2 \equiv \emptyset$ in (16), we derive Cover-Leung achievable rate region [6] for MAC with feedback.

*Achievability:* The proof follows the same lines as that of Theorem 1 except that in this case we have no encoding error. Therefore by negligibly altering the codebook generated we can easily derive the expressions in (16). In fact, if the senders want to send $(m_{1b}, m_{2b})$ to the receiver, upon choosing the helping index $m_{0b}$, they find the suitable $u^n, v_1^n$ and $v_2^n$ sequences directly and then send $x_1^n$ and $x_2^n$ where

$$x_{ki} = f_k(u_i(m_0), v_{ki}(m_{0b}, m_{kb}), s_0^i, s_k^i), \quad k = 1,2$$

Notice that during the codebook generation process, instead of generating $2^{n(R_k+R_k')}, k = 1,2$, n-sequences $v_k^n$, we only generate $2^{nR_k}$ such sequences. We omit the details for brevity.

*Converse:* Suppose that there exists a code $((2^{nR_1}, 2^{nR_2}), n)$ with $P_e^{(n)} \to 0$ as $n \to \infty$ for the discrete memoryless MAC with two-sided feedback and with SW-type correlated SI causally available at the transmitters. According to Fano's inequality we have:

$$nR_1 = H(M_1) = H(M_1|M_0M_2S_0^nS_2^n)$$
$$= H(M_1|M_0M_2S_0^nS_2^nY^n) + I(M_1;Y^n|M_0M_2S_0^nS_2^n)$$
$$\leq n\epsilon_{1n} + \sum_{i=1}^n I(M_1;Y_i|M_0M_2S_0^nS_2^nY^{i-1})$$
$$= n\epsilon_{1n} + \sum_{i=1}^n H(Y_i|M_0M_2S_0^nS_2^nY^{i-1})$$
$$- \sum_{i=1}^n H(Y_i|M_0M_1M_2S_0^nS_2^nY^{i-1})$$
$$\leq n\epsilon_{1n} + \sum_{i=1}^n H(Y_i|M_0M_2S_0^iS_2^iY^{i-1})$$
$$- \sum_{i=1}^n H(Y_i|M_0M_1M_2S_0^nS_1^nS_2^nY^{i-1})$$
$$\overset{(a)}{=} n\epsilon_{1n} + \sum_{i=1}^n H(Y_i|M_0M_2S_0^iS_2^iY^{i-1})$$
$$- \sum_{i=1}^n H(Y_i|M_0M_1M_2S_0^nS_1^nS_2^nY^{i-1}X_{1i}X_{2i})$$
$$\overset{(b)}{=} n\epsilon_{1n} + \sum_{i=1}^n H(Y_i|M_0M_2S_0^iS_2^iY^{i-1})$$
$$- \sum_{i=1}^n H(Y_i|M_0M_1M_2S_0^iS_1^iS_2^iY^{i-1}X_{1i}X_{2i})$$
$$\overset{(c)}{=} n\epsilon_{1n} + \sum_{i=1}^n H(Y_i|M_0M_2S_0^iS_2^iY^{i-1})$$
$$- \sum_{i=1}^n H(Y_i|M_0M_1M_2S_0^iS_1^iS_2^iY^{i-1})$$
$$= n\epsilon_{1n} + \sum_{i=1}^n H(Y_i|M_0M_2S_0^iS_2^iY^{i-1})$$
$$- \sum_{i=1}^n H(Y_i|M_0M_1M_2S_0^iS_1^iS_2^iY^{i-1})$$
$$= n\epsilon_{1n} + \sum_{i=1}^n I(M_0M_1S_0^iS_1^iY^{i-1};Y_i|M_0M_2S_0^iS_2^iY^{i-1})$$
$$= n\epsilon_{1n} + \sum_{i=1}^n I(V_{1i};Y_i|U_iV_{2i}), \qquad (18)$$

where

$$U_i \triangleq (S_0^i, Y^{i-1})$$
$$V_{1i} = (M_0M_1S_0^iS_1^iY^{i-1})$$
$$V_{2i} = (M_0M_2S_0^iS_2^iY^{i-1})$$

Now $(a)$ and $(c)$ follow from

$$x_{1i} = f_1(u_i(M_0), v_{1i}(M_0, M_1), s_0^i, s_1^i) = f_1(M_0M_1s_0^is_1^i)$$
$$x_{2i} = f_2(u_i(M_0), v_{2i}(M_0, M_2), s_0^i, s_2^i) = f_2(M_0M_2s_0^is_2^i)$$ (19)

and $(b)$ follows from the memorylessness of the channel, i.e. the following Markov chain

$$(M_0M_1M_2S_0^{i-1}S_{0,i+1}^nS_1^{i-1}S_{1,i+1}^nS_2^{i-1}S_{2,i+1}^nY^{i-1}) \to$$
$$(X_{1i}X_{2i}S_{0i}S_{1i}S_{2i}) \to Y_i$$ (20)

Notice that according to the variables defined for $U_i, V_{1i},$ and $V_{2i}$, (18) can also be written as

$$nR_1 = n\epsilon_{1n} + \sum_{i=1}^n I(V_{1i};Y_i|U_iV_{2i}S_{0i}S_{2i}).$$

Bound on $R_2$ is derived exactly the same as that of $R_1$.

$$n(R_1 + R_2) = H(M_1M_2|Y^n) + I(M_1M_2;Y^n)$$
$$\leq n\epsilon_{12n} + \sum_{i=1}^n I(M_1M_2;Y_i|Y^{i-1})$$
$$\leq n\epsilon_{12n} + \sum_{i=1}^n I(M_0M_1M_2S_0^iS_1^iS_2^iY^{i-1};Y_i)$$
$$= n\epsilon_{12n} + \sum_{i=1}^n I(V_{1i}V_{2i};Y_i). \qquad \blacksquare$$

Define $C_{MACTSF-SCSI}(p)$ as the set of all rate pairs $(R_1, R_2)$ such that

$$R_1 \leq I(V_1;Y|UV_2)$$
$$R_1 \leq I(V_1;Y|UV_2) \qquad (21)$$
$$R_1 + R_2 \leq I(V_1V_2;Y)$$

for some $p \in \mathcal{P}_1^*$.

***Theorem 3:*** The set $C_{MACTSF-SCSI}^*$ is the capacity region of the discrete memoryless MAC with two-sided feedback and with correlated SI strictly causally available at the transmitters

$$C_{MACTSF-SCSI}^* = cl\left(co\left(\bigcup_{p \in \mathcal{P}_1^*} C_{MACTSF-SCSI}(p)\right)\right) \qquad (22)$$

***Achievability:*** The proof follows the same lines as those of Theorem 2 with a negligible variation, i.e. here we have

$$x_{ki} = f_k(u_i(m_0), v_{ki}(m_{0b}, m_{kb}), s_0^{i-r}, s_k^{i-r}) \quad k = 1,2$$

Notice that in this case, the encoders do not have non-causal access to the n-sequence of their state information at the end of each block.

*Converse:* Suppose that there exists a code $((2^{nR_1}, 2^{nR_2}), n)$ with $P_e^{(n)} \to 0$ as $n \to \infty$ for the discrete memoryless MAC with two-sided feedback and with SW-type correlated SI strictly causally available at the transmitters. According to Fano's inequality for some $0 < r < i$ we have

$$nR_1 = H(M_1|M_0M_2S_0^nS_2^nY^n) + I(M_1;Y^n|M_0M_2S_0^nS_2^n)$$

$$\leq n\epsilon_{1n} + \sum_{i=1}^{n} I(M_1;Y_i|M_0M_2S_0^nS_2^nY^{i-1})$$

$$= n\epsilon_{1n} + \sum_{i=1}^{n} H(Y_i|M_0M_2S_0^nS_2^nY^{i-1})$$

$$- \sum_{i=1}^{n} H(Y_i|M_0M_1M_2S_0^nS_2^nY^{i-1})$$

$$\leq n\epsilon_{1n} + \sum_{i=1}^{n} H(Y_i|M_0M_2S_0^{i-r}S_2^{i-r}S_{0i}S_{2i}Y^{i-1})$$

$$- \sum_{i=1}^{n} H(Y_i|M_0M_1M_2S_0^nS_1^nS_2^nY^{i-1}X_{1i}X_{2i})$$

$$\stackrel{(a)}{=} n\epsilon_{1n} + \sum_{i=1}^{n} H(Y_i|M_0M_2S_0^{i-r}S_2^{i-r}S_{0i}S_{2i}Y^{i-1})$$

$$- \sum_{i=1}^{n} H(Y_i|M_0M_1M_2S_0^{i-r}S_1^{i-r}S_2^{i-r}S_{0i}S_{1i}S_{2i}Y^{i-1}X_{1i}X_{2i})$$

$$\stackrel{(b)}{\leq} n\epsilon_{1n} + \sum_{i=1}^{n} H(Y_i|M_0M_2S_0^{i-r}S_2^{i-r}S_{0i}S_{2i}Y^{i-1})$$

$$- \sum_{i=1}^{n} H(Y_i|M_0M_1M_2S_0^{i-r}S_1^{i-r}S_2^{i-r}S_{0i}S_{1i}S_{2i}Y^{i-1}) = n\epsilon_{1n}$$

$$+ \sum_{i=1}^{n} I(M_0M_1S_0^{i-r}S_1^{i-r}S_{0i}S_{1i}Y^{i-1};Y_i|M_0M_2S_0^{i-r}S_2^{i-r}S_{0i}S_{2i}Y^{i-1})$$

$$= n\epsilon_{1n} + \sum_{i=1}^{n} I(V_{1i};Y_i|U_iV_{2i}),$$

where $(a)$ follows from Markov chain (20) and $(b)$ follows from the strict causal version of (19), i.e.

$$x_{1i} = f_1(M_0M_1s_0^{i-r}s_1^{i-r})$$
$$x_{2i} = f_2(M_0M_2s_0^{i-r}s_2^{i-r})$$

and

$$U_i \triangleq (S_0^{i-r}, S_{0i}, Y^{i-1})$$
$$V_{1i} \triangleq (M_0M_1S_0^{i-r}S_1^{i-r}S_{0i}S_{1i}Y^{i-1})$$
$$V_{2i} \triangleq (M_0M_2S_0^{i-r}S_2^{i-r}S_{0i}S_{2i}Y^{i-1})$$

Bounds on $R_2$ is the same and bound on $R_1 + R_2$ is straightforward and is omitted for brevity. ∎

Now consider the MAC with partial feedback and with correlated SI non-causally available at the transmitters depicted in Fig. 4. We provide an achievable rate region for this type of channel. For this case, we use the notions of superposition, Gelf'and-Pinsker, and block Markov encoding and deterministic partitioning, and *restricted decoding*.

Define $\mathcal{R}_{MACPF-NCSI}(p)$ as the set of all rate pairs $(R_1, R_2)$ satisfying

$$\begin{aligned}R_1 &\leq I(V_1;Y|UV_2) - I(V_1;S_0S_1|U) \\ R_2 &\leq \min\{I(V_2;Y|UV_1) + I(U;Y), I(V_2;Y|UV_1S_0S_1)\} \\ &\quad - I(V_2;S_0S_2|U) \\ R_1 + R_2 &\leq I(V_1V_2;Y) - I(V_2;S_0S_2|U) - I(V_1;S_0S_1|U)\end{aligned} \quad (23)$$

for some $p \in \mathcal{P}_1^*$.

*Theorem 4:* The set $\mathcal{R}_{MACPF-NCSI}^*$ is achievable for the discrete memoryless MAC with partial feedback and with correlated SI non-causally available at the transmitters

$$\mathcal{R}_{MACPF-NCSI}^* = cl\left(co\left(\bigcup_{p \in \mathcal{P}_1^*} \mathcal{R}_{MACPF-NCSI}(p)\right)\right) \quad (24)$$

*Codebook Generation:* Assume that

$$p(s_0^n, s_1^n, s_2^n) = \prod_{i=1}^{n} p(s_{0i}) p(s_{1i}) p(s_{2i})$$

Fix a joint distribution as $p(u)p(v_1|u, s_0, s_1)p(v_2|u, s_0, s_2)$. Randomly and independently generate $2^{nR_0}$ sequences $u^n(m_0)$, and conditionally independently superimpose $2^{n(R_1'+R_1)}$, and $2^{n(R_2'+R_2)}$ n-sequences of $v_1^n(m_0, m_1', m_1)$ and $v_2^n(m_0, m_2', m_2)$ as in Theorem 1 and randomly pour them into $2^{nR_1}$ and $2^{nR_2}$ bins, respectively. Call these bins the Gelf'and-Pinsker (GP) bins. Now partition the $2^{nR_2}$ GP-bins into $M_0 = 2^{nR_0}$ partitions in a deterministic way and call the latter partitions D-partitions. Set $c_{m_2}$ as the number of the cell inside which $m_2$ is located with $c_{m_2} \in \{1, 2, \ldots, M_0\}$.

*Encoding:* In here, just like Theorem 1, $B$ blocks of communication each of length $n$ are used.

Now assume that the messages $m_{1,b}$ and $m_{2,b}$ are to be sent to the receiver in block $b$ and also suppose that the receiver has knowledge of $\left((\hat{m}_{1,1}, \hat{m}_{2,1}), (\hat{m}_{1,2}, \hat{m}_{2,2}), \ldots, (\hat{m}_{1,b-2}, \hat{m}_{2,b-2})\right)$ and the first transmitter has knowledge of $(\hat{\hat{m}}_{2,1}, \hat{\hat{m}}_{2,2}, \ldots, \hat{\hat{m}}_{2,b-1})$ at the *beginning* of block $b$. We want to make sure that the receiver can estimate $(\hat{m}_{1,b-1}, \hat{m}_{2,b-1})$ and the first sender can estimate $\hat{\hat{m}}_{2,b}$ with low probability of error *at the end of* block $b$.

Assume that the transmitters desire to send $(m_{1,b}, m_{2,b})$ at the beginning of block $b$. The first sender searches in GP bin $m_{1,b}$

and looks for some $m'_{1,b}$ such that

$$\left(u^n(\widehat{\widehat{m}}_{0,b}), v_1^n(\widehat{\widehat{m}}_{0,b}, m'_{1,b}, m_{1,b}), s_0^n, s_1^n\right) \in A_\epsilon^{(n)}$$

where $c_{\widehat{m}_{2,b-1}} = \widehat{\widehat{m}}_{0,b}$. The index is reliably found if

$$R'_1 \geq I(V_1; S_0 S_1 | U) \quad (25)$$

Let the index chosen be $m'_{1,b} = M'_{1,b}$.

The second sender searches in GP bin $m_{2,b}$ and looks for some $m'_{2,b}$ such that

$$\left(u^n(m_{0,b}), v_2^n(m_{0,b}, m'_{2,b}, m_{2,b}), s_0^n, s_2^n\right) \in A_\epsilon^{(n)}$$

where $c_{m_{2,b-1}} = m_{0,b}$. The index $m'_{2,b}$ is found with negligible error probability if

$$R'_2 \geq I(V_2; S_0 S_2 | U) \quad (26)$$

Let the index chosen be $m'_{2,b} = M'_{2,b}$.

The transmitters send $x_1^n(b)$ and $x_2^n(b)$ where

$$x_{1i}(b) = f_1\big(u_i(\widehat{\widehat{m}}_{0,b}), v_{1i}(\widehat{\widehat{m}}_{0,b}, M'_{1,b}, m_{1,b}), s_0^n, s_1^n\big)$$
$$x_{2i}(b) = f_2\big(u_i(m_{0,b}), v_{2i}(m_{0,b}, M'_{2,b}, m_{2,b}), s_0^n, s_2^n\big)$$

The receiver receives $y^n(b)$ and declares $\widehat{m}_{0,b} = m_{0,b}$ to be the index sent if

$$\left(u^n(m_{0,b}), y^n(b)\right) \in A_\epsilon^{(n)}$$

This stage is accomplished if

$$R_0 \leq I(U; Y) \quad (27)$$

The first sender, meanwhile, receives $y^n(b)$ through the feedback link and declares $\widehat{\widehat{m}}_{2,b} = m_{2,b}$ as the message sent by the second transmitter in block $b$ if

$$(u^n(\widehat{\widehat{m}}_{0,b}), v_1^n(\widehat{\widehat{m}}_{0,b}, M'_{1,b}, m_{1,b}), v_2^n(m_{0,b}, \widehat{\widehat{M}}'_{2,b}, \widehat{\widehat{m}}_{2,b})$$
$$, s_0^n, s_1^n, y^n(b)) \in A_\epsilon^{(n)})$$

This stage is accomplished with small probability of error if

$$R_2 + R'_2 \leq I(V_2; Y | UV_1 S_0 S_1) \quad (28)$$

Now the decoder declares $(\widehat{m}_{1,b-1}, \widehat{m}_{2,b-1})$ as the messages sent in block $b-1$ if

$$(u^n(\widehat{m}_{0,b-1}), v_1^n(\widehat{m}_{0,b-1}, \widehat{m}'_{1,b-1}, \widehat{m}_{1,b-1})$$
$$, v_2^n(\widehat{m}_{0,b-1}, \widehat{m}'_{2,b-1}, \widehat{m}_{2,b-1}), y^n(b-1)) \in A_\epsilon^{(n)})$$

where $c_{\widehat{m}_{2,b-2}} = \widehat{m}_{0,b-1}$ with $\widehat{m}_{0,b-1}$ previously known to the receiver in a stage like (27). Here we see that the message $m_{02}$ is restricted to stay in a specific cell thus limiting the number of possible messages.

The messages are reliably decoded provided that

$$R_1 + R'_1 + R_2 + R'_2 - R_0 \leq I(V_1 V_2; Y | U) \quad (29)$$
$$R_2 + R'_2 - R_0 \leq I(V_2; Y | UV_1) \quad (30)$$
$$R_1 + R'_1 \leq I(V_1; Y | UV_2) \quad (31)$$

Combining (28), (29), (30), and (31) with (25), (26), and (27) we derive (23). ■

We now give an achievable rate region for the capacity of the discrete memoryless MAC with partial feedback and with correlated SI causally available at the transmitters.

Define $C_{MACPF-CSI}(p)$ as the set of all rate pairs $(R_1, R_2)$ satisfying

$$R_1 \leq I(V_1; Y | UV_2)$$
$$R_2 \leq \min\{I(V_2; Y | UV_1) + I(U; Y), I(V_2; Y | UV_1 S_0 S_1)\} \quad (32)$$
$$R_1 + R_2 \leq I(V_1 V_2; Y)$$

for some $p \in \mathcal{P}_1^*$.

***Theorem 5:*** The set $C^*_{MACPF-CSI}$ is the capacity region of the discrete memoryless MAC with partial feedback and with SW-type correlated SI causally available at the transmitters

$$C^*_{MACPF-CSI} = cl\left(co\left(\bigcup_{p \in \mathcal{P}_1^*} C_{MACPF-CSI}(p)\right)\right) \quad (33)$$

***Achievability:*** The proof of achievability follows the same lines as those of Theorem 4 except that in here we do not have any encoding error and the codebook is generated with a slight variation than to Theorem 4.

***Converse:*** Suppose that there exists a code $\left((2^{nR_1}, 2^{nR_2}), n\right)$ with $P_e^{(n)} \to 0$ and $n \to \infty$ for the discrete memoryless MAC with partial feedback and with SW-type correlated SI causally available at the transmitters. According to Fano's inequality we have

$$nR_1 \leq n\epsilon_{1n} + \sum_{i=1}^n I(M_1; Y_i | M_0 M_2 S_0^n S_2^n Y^{i-1})$$
$$= n\epsilon_{1n} + \sum_{i=1}^n H(Y_i | M_0 M_2 S_0^n S_2^n Y^{i-1})$$
$$\quad - \sum_{i=1}^n H(Y_i | M_0 M_1 M_2 S_0^n S_2^n Y^{i-1})$$
$$\leq n\epsilon_{1n} + \sum_{i=1}^n H(Y_i | M_0 M_2 S_0^i S_2^i)$$
$$\quad - \sum_{i=1}^n H(Y_i | M_0 M_1 M_2 S_0^n S_1^n S_2^n Y^{i-1} X_{1i} X_{2i})$$
$$\overset{(a)}{=} n\epsilon_{1n} + \sum_{i=1}^n H(Y_i | M_0 M_2 S_0^i S_2^i)$$

$$-\sum_{i=1}^{n} H(Y_i|M_0M_1M_2S_0^iS_1^iS_2^iY^{i-1}X_{1i}X_{2i})$$

$$\stackrel{(b)}{=} n\epsilon_{1n} + \sum_{i=1}^{n} H(Y_i|M_0M_2S_0^iS_2^i)$$

$$-\sum_{i=1}^{n} H(Y_i|M_0M_1M_2S_0^iS_1^iS_2^i)$$

$$= n\epsilon_{1n} + \sum_{i=1}^{n} I(M_0M_1S_0^iS_1^i; Y_i|M_0M_2S_0^iS_2^i)$$

$$\leq n\epsilon_{1n} + \sum_{i=1}^{n} I(M_0M_1S_0^iS_1^iY^{i-1}; Y_i|M_0M_2S_0^iS_2^i)$$

$$= n\epsilon_{1n} + \sum_{i=1}^{n} I(V_{1i}; Y_i|U_iV_{2i}),$$

where $(a)$ follows from Markov chain (23) and $(b)$ follows from (22) and

$$U_i \triangleq (M_0, S_0^i)$$
$$V_{1i} \triangleq (M_0, M_1, S_0^i, S_1^i, Y^{i-1})$$
$$V_{2i} \triangleq (M_0, M_2, S_0^i, S_2^i)$$

For the bound on $R_2$ we have

$$nR_2 \leq n\epsilon_{2n} + \sum_{i=1}^{n} I(M_2; Y_i|M_0M_1S_0^nS_1^nY^{i-1})$$

$$= n\epsilon_{2n} + \sum_{i=1}^{n} H(Y_i|M_0M_1S_0^nS_1^nY^{i-1})$$

$$-\sum_{i=1}^{n} H(Y_i|M_0M_1M_2S_0^nS_1^nY^{i-1})$$

$$\stackrel{(a)}{=} n\epsilon_{2n} + \sum_{i=1}^{n} H(Y_i|M_0M_1S_0^nS_1^nY^{i-1})$$

$$-\sum_{i=1}^{n} H(Y_i|M_0M_1M_2S_0^nS_1^nY^{i-1}X_{1i}X_{2i})$$

$$\stackrel{(b)}{=} n\epsilon_{2n} + \sum_{i=1}^{n} H(Y_i|M_0M_1S_0^nS_1^nY^{i-1})$$

$$-\sum_{i=1}^{n} H(Y_i|M_0M_1M_2S_0^iS_1^iY^{i-1}X_{1i}X_{2i})$$

$$\stackrel{(c)}{\leq} n\epsilon_{2n} + \sum_{i=1}^{n} H(Y_i|M_0M_1S_0^iS_1^iY^{i-1})$$

$$-\sum_{i=1}^{n} H(Y_i|M_0M_1M_2S_0^iS_1^iY^{i-1})$$

$$= n\epsilon_{2n} + \sum_{i=1}^{n} I(M_0M_2S_0^i; Y_i|M_0M_1S_0^iS_1^iY^{i-1})$$

$$\leq n\epsilon_{2n} + \sum_{i=1}^{n} I(M_0M_2S_0^iS_2^i; Y_i|M_0M_1S_0^iS_1^iY^{i-1})$$

$$= n\epsilon_{2n} + \sum_{i=1}^{n} I(V_{2i}; Y_i|U_iV_{1i})$$

$$\leq n\epsilon_{2n} + \sum_{i=1}^{n} I(V_{2i}; Y_i|U_iV_{1i}) + I(M_0S_0^i; Y_i)$$

$$= n\epsilon_{2n} + \sum_{i=1}^{n} I(V_{2i}; Y_i|U_iV_{1i}) + I(U_i; Y_i),$$

where $(a)$ follows from (19), $(b)$ follows from (20), and $(c)$ follows from (19) and the fact that removing conditioning increases the entropy.

The other bound on $nR_2$ is

$$nR_2 \leq n\epsilon_{2n} + \sum_{i=1}^{n} I(M_0M_2S_0^iS_2^i; Y_i|M_0M_1S_0^iS_1^iY^{i-1})$$

$$= n\epsilon_{2n} + \sum_{i=1}^{n} I(V_{2i}; Y_i|U_iV_{1i}S_{0i}S_{1i})$$

Bound on $n(R_1 + R_2)$ is straightforward and is omitted. ∎

*Remark:* Notice that in this converse proof, the definition of the auxiliary random variables is such that we see $Y^{i-1}$ in $V_{1i}$ but there is no $Y^{i-1}$ in $V_{2i}$ and that is because only the first transmitter is fed back from the output of the channel and therefore only the first transmitter uses the feedback information in constructing the auxiliary random variable.

Define $C_{MACPF-SCSI}(p)$ as the set of all rate pairs $(R_1, R_2)$ satisfying

$$\begin{aligned} R_1 &\leq I(V_1; Y|UV_2) \\ R_2 &\leq I(V_2; Y|UV_1) \\ R_1 + R_2 &\leq I(V_1V_2; Y) \end{aligned} \quad (34)$$

for some $p \in \mathcal{P}_1^*$.

*Theorem 6:* The set $C^*_{MACPF-SCSI}$ is achievable for the discrete memoryless MAC with partial feedback and with correlated SI strictly causally available at the transmitters

$$C^*_{MACPF-SCSI} = cl\left(co\left(\bigcup_{p \in \mathcal{P}_1^*} C_{MACPF-SCSI}(p)\right)\right) \quad (35)$$

*Proof:* The proof follows approximately the same lines as that of Theorem 5 and is omitted. ∎

## Conclusion

Achievable rate and capacity regions for the Multiple Access Channel with full and partial feedback with SW-type correlated side information available non-causally, causally, and strictly causally at the transmitters were provided. Cover-Leung's achievable rate was shown to be a special case of our region for discrete memoryless MAC with full feedback and with correlated SI.

## References

[1] R. Ahlswede, "Multi-way Communication Channels," in *Proc. Second Int. Symp. Inform. Transmiss.*, Tsahkadsor, Armenia, USSR, Hungarian Press, 1971.

[2] H. Liao, "A coding theorem for multiple access communications," presented at the *Int. Symp. Inform. Theory,* Asimolar, 1972; Also Ph.D. dissertation, "Multiple access channels," Dept. Electrical Engineering, Univ. Hawaii, 1972.

[3] D. Slepian and J. K. Wolf, "A coding theorem for multiple access channels with correlated sources,*" Bell Syst. Tech. J.,* vol. 52, pp. 1037-1076, Sept. 1973.

[4] T. Cover, A. El Gamal, M. Salehi, "Multiple access channels with arbitrarily correlated sources," *IEEE Trans. Inform. Theory,* vol. 26, pp. 648-657, Nov. 1980.

[5] N. T. Gaarder and J. K. Wolf, "The capacity region of a multiple access discrete memoryless channel can increase with feedback," *IEEE Trans. Inform. Theory*, vol. 21, pp. 100-102, Jan.1975.

[6] T. Cover and C. S. K. Leung, "An achievable rate region for the multiple-access channel with feedback," *IEEE Trans. Inform. Theory,* vol. 27, pp. 292-298, May. 1981.

[7] Frans M. J. Willems, "The feedback capacity region of a class of discrete memoryless multiple access channels," *IEEE Trans. Inform. Theory,* vol. 28, pp. 93-95, Jan. 1982.

[8] Frans. M. J. Willems and E. C. Van Der Meulen, "Partial feedback for the discrete memoryless multiple access channel," *IEEE Trans. Inform. Theory*, vol. 29, pp. 287-290, March 1983.

[9] C. E. Shannon, "Channels with side information at the transmitter," *IBM Journal Res. And Develop,* vol. 2, pp. 289-293, Oct. 1958.

[10] S. I. Gelf'and and M. S. Pinsker, "Coding for channel with random paratmeters," *Probl. of Con. Inform. Theory,* vol. 52, no. 12, pp. 19–31, Jan. 1980.

[11] S. Sigurjonsson and Y. H. Kim, "On multiple user channels with causal state informationat the transmitters," in *Proc. IEEE Int. Symp. Inform. Theory*, Adelaide, Australia, pp. 72-76, Sept. 2005.

[12] A. A. Jafar, "Capacity with causal and non-causal side information: A unified view," *IEEE Trans. Inform. Theory*, vol. 52, pp. 5468–5474, Dec. 2006.

[13] T. Philosof, R. Zamir, U. Erez, "Technical Report: Achievable rates for the MAC with correlated channel-state information," Dept. of Electrical Engineering – systems, Tel-Aviv University, Technical Report, Dec. 2008.

[14] Amos Lapidoth and Yossef Steinberg, "The multiple access channel with causal and non-causal side information at the encoders," *Int. Zurich Seminar on Communications (IZS),* March 3-5, 2010.

[15] R. Khosravi-Farsani and F. Marvasti, "Capacity bounds for multiuser channels with non-causal channel state information at the transmitters," *Infotm. Theory Workshop (ITW)*, pp. 195-199, 16-20 Oct. 2011.

[16] Elham Bahmani and Ghosheh Abed Hodtani, "Achievable rate regions for the dirty maultiple access channel with partial side information at the transmitters," in *Proc. IEEE Int. Symp. Inform. Theory*, pp. 1707-1711, Cambridge, MA, USA, 1-6 July 2012.

[17] R. Khosravi-Farsani and F. Marvasti, "Multiple access channels with cooperative encoders and channel state information," accepted for publication in *Transactions on Emerging Telecommunications Technologies.*